\documentclass[twocolumn]{jpsj3}
\usepackage{txfonts}
\usepackage{color}

\title{Thermodynamic Investigation of Metamagnetic Transitions and Partial Disorder
\\ in the Quasi-Kagome Kondo Lattice CePdAl}

\author{Kenji Mochidzuki\thanks{kmochi@issp.u-tokyo.ac.jp}$^{1}$, 
Yusei Shimizu\thanks{yuseishimizu@imr.tohoku.ac.jp, Present address: Institute for Materials Research, Tohoku University, Oarai, 311-1313, Japan}$^{1}$,
Akihiro Kondo$^{1}$, 
Shota Nakamura$^{1}$, 
Shunichiro Kittaka$^{1}$, Yohei Kono$^{1}$, \\ 
Toshiro Sakakibara$^{1}$, Yoichi Ikeda$^2$, 
Yosikazu Isikawa$^3$, 
and 
Koichi Kindo$^{1}$
}
\inst{$^{1}$Institute for Solid State Physics, the University of Tokyo, Kashiwa, 277-8581, Japan \\
$^{2}$Institute for Materials Research, Tohoku University, Sendai, 980-0811, Japan\\
$^{3}$Graduate School of Science and Engineering, University of Toyama, Toyama, 930-8555, Japan
}
\abst{
Low-temperature ($T$) thermodynamic properties in magnetic fields ($B$) of the quasi-Kagome antiferromagnet CePdAl ($T_{\rm N}\sim$ 2.7 K) were studied by means of magnetization and specific-heat $C(T,B)$ measurements. 
The unusual magnetic phase diagram, including three thermodynamic phases I-III, and crossover anomalies, has been obtained. 
In low-field phase I, the existence of the partial Kondo screening state on the one-third Ce sites is supported from the thermodynamic point of view: (i) an appearance of the paramagnetic moment with one-third  of the full moment around 3 T, and (ii) an occurrence of the entropy release $\sim$0.3$R$ln2 around the observed crossover.
On the other hand, a strong enhancement of $C/T$ is found around the point, where $T_{\rm N}(B)$ meets the boundary of phase III, indicating the possible presence of a tri-critical point.
Furthermore, it is suggested that the three metamagnetic transitions at $B_{{\rm m}i}$ ($i=$1,2,3) come from  successive increments of the Ising-type magnetic moment along the $c$-axis. 
 In high-field state above 20 T, the behavior of $f$ electron can be rather described by the crystal-electric-field model.
}
\begin{document}
\maketitle

%%%%%%%%%%%%%%%%%%%%%%%%%%%%%%%%%%%%%
\section{\label{sec:level1}Introduction}
%%%%%%%%%%%%%%%%%%%%%%%%%%%%%%%%%%%%%

The geometrical frustration in magnetic systems often causes attractive phenomena such as the spin liquid\cite{Lee2008,Leon2010}, spin ice\cite{Bramwell2001}, and spin nematic order\cite{Momoi2005}. Over several decades, a number of studies have been performed on insulator quantum-spin systems. Recently these interests have been extended to strongly correlated systems with conduction electrons\cite{Intro_frustrate}. In such systems, the quantum mechanical effects between conduction electrons and localized spin with frustration can play an important role to lift the frustration. To explore novel quantum phases, it is intriguing to study geometrical frustration effects on strongly-correlated $f$-electron heavy-fermion systems, such as UNi$_4$B\cite{Mentink1994,Lacroix1996} and CePdAl\cite{Donni1996,Keller2002}. 
 
%%%%%%%%%%%%%%%%%%%%%%%%%%%%%%%%%%%%%%%%%%%%%%%%%%%%%%%%%%%%%%%%%%%%%%%%%%%%%%%%%%%%%%%%%%

In the present study, we focus on the geometrically frustrated Kondo lattice CePdAl, which crystallizes in a hexagonal ZrNiAl-type crystal structure (space group $P\bar{6}2m$) with no inversion symmetry\cite{Xue1994}. There exist three equivalent Ce sites, forming a quasi-Kagome lattice\cite{Hulliger1993,Schank1994}. The magnetic susceptibility has a strong anisotropy due to the crystal-electric-field (CEF) effect\cite{Isikawa1996}. An incommensurate antiferromagnetic order occurs at $T_{\rm{N}}$ = 2.7 K\cite{Kitazawa1994}. The smallness of energy scale of $T_{\rm{N}}$ compared to the paramagnetic Curie temperature of $\theta_{\rm{p}} = -$34 K\cite{Donni1996} may indicate the presence of a strong frustration in this system.

%%%%%%%%%%%%%%%%%%%%%%%%%%%%%%%%%%%%%%%%%%%%%%%%%%%%%%%%%%%%%%%%%%%%%%%%%%%%%%%%%%%%%%%%%%

One of the most interesting properties in CePdAl is the presence of partially disordered spins due to the Kondo effect, coexistent with the antiferromagnetic order below $T_{\rm{N}}$. The magnetic structure below $T_{\rm{N}}$ is characterized by an incommensurate propagation vector $Q$ = (0.5, 0, $\tau$)\cite{Donni1996}, where the component $\tau$ decreases with cooling and becomes constant ($\tau\approx$ 0.354) below $\sim$2 K\cite{Keller2002,Prokes2006}. In the hexagonal basal plane, the ordered moments at two-thirds of Ce sites, \textit{i.e.}, Ce(1) and Ce(3), form ferromagnetic chains, which couple antiferromagnetically each other\cite{Donni1996}. On the other hand, surprisingly, one-third of Ce sites, \textit{i.e.}, Ce(2), which are located between the magnetic chains formed by Ce(1) and Ce(3), have no magnetic order below $T_{\rm{N}}$\cite{Donni1996}.  Such partially disordered behavior has also been revealed from NMR measurements; the spin-lattice relaxation time obeys Korringa's law down to 30 mK without further magnetic transition below $T_{\rm{N}}$\cite{Oyamada2008}. These experimental facts indicate that the moments on the nonmagnetic Ce(2) sites are screened by the Kondo effect, and the frustration is lifted by forming the heavy-fermion state. Indeed, the low-temperature electronic specific-heat coefficient becomes huge $C/T \sim$ 1 JK$^{-2}$mol$^{-1}$ just above $T_{\rm{N}}$, and the estimated Kondo temperature has a similar energy scale $T_{\mathrm{K}}\sim$ 6 K\cite{Cermak_JPhysC_2010} to $T_{\rm{N}}$.

%%%%%%%%%%%%%%%%%%%%%%%%%%%%%%%%%%%%%%%%%%%%%%%%%%%%%%%%%%%%%%%%%%%%%%%%%%%%%%%%%%%%%%%%%%

In magnetic fields, the magnetization exhibits three successive metamagnetic transitions at 0.51 K below 5 T and reaches 1.43 $\mu_{\rm B}$/Ce at 10 T\cite{Hane2000}, smaller than the value of 1.58-1.81 $\mu_{\rm B}$/Ce observed in the neutron diffraction  experiments\cite{Donni1996,Prokes2015}. Thus experiments in further high magnetic fields are necessary to understand the whole magnetization process. Moreover, recently, a rich $B$-$T$ phase diagram with several field-induced states has been revealed\cite{Zhao2016}. However, the details of the ground states, their anisotropy, and effect of magnetic fluctuation in the high magnetic phases have still not been clarified. Therefore thermodynamic understanding of the field-induced magnetic phases in CePdAl and their relationship to the heavy-fermion state is necessary by means of precise quantitative measurements.

%%%%%%%%%%%%%%%%%%%%%%%%%%%%%%%%%%%%%%%%%%%%%%%%%%%%%%%%%%%%%%%%%%%%%%%%%%%%%%%%%%%%%%%%%%

In this paper, we report the results of the high-field magnetization and specific-heat measurements on CePdAl. To reveal the magnetic property above the metamagnetic transitions and the magnetic anisotropy in the high-field phases, we have measured the magnetization curves from the magnetic easy axis ($c$-axis) to the hard plane ($ab$-plane) by using the pulsed magnetic field up to 50 T. From the specific-heat measurements for $B\parallel c$-axis up to 7 T, we have also precisely constructed magnetic phase diagram of low-$T$ ground state of CePdAl and investigated the effect of magnetic fluctuation in the high-field phases.

%%%%%%%%%%%%%%%%%%%%%%%%%%%%%%%%%%
\section{\label{sec:level2}Experiment}
%%%%%%%%%%%%%%%%%%%%%%%%%%%%%%%%%%

A single crystalline sample of CePdAl was prepared by the Czochralski pulling method\cite{Isikawa1996}. The sample was orientated by the Laue X-ray photographs. The high magnetic field up to 50 T for the magnetization measurement was generated by using a non-destructive pulsed magnet, and the sample was cooled by using a $^4$He cryostat ($T\geq$ 1.3 K, $B\leq$ 50 T). The magnetization was measured by a standard pick-up coil method. To apply magnetic fields along various crystal orientations, the sample was put between diagonally cut quartz rods and sealed into a heat shrinkable tube. Then, we could avoid the effect of the torque caused by the large magnetic anisotropy. The specific-heat measurements were performed by a standard quasi-adiabatic heat-pulse technique with a hand-made calorimeter installed in a $^3$He Oxford Heliox system ($T\geq$ 0.3 K, $B\leq$ 7 T).  

%%%%%%%%%%%%%%%%%%%%%%%%%%%%%%%%%%%%%%%%%%%%
\section{\label{sec:level3}Results and discussion}
%%%%%%%%%%%%%%%%%%%%%%%%%%%%%%%%%%%%%%%%%%%%

%\subsection{\label{sec:level30}Magnetization}

%%%%%%%%%%%%%%%%%%%%%%%%%%%%%%%%%%%%%%%%%%%%%%%%%%%%%%%%%%%%%%%%%%%%%%%%%%%%%%%%%%%%%%%%%%%%%%%%%%%%%%%%%%%%
\begin{figure}[t]
\includegraphics[width = 3.4in]{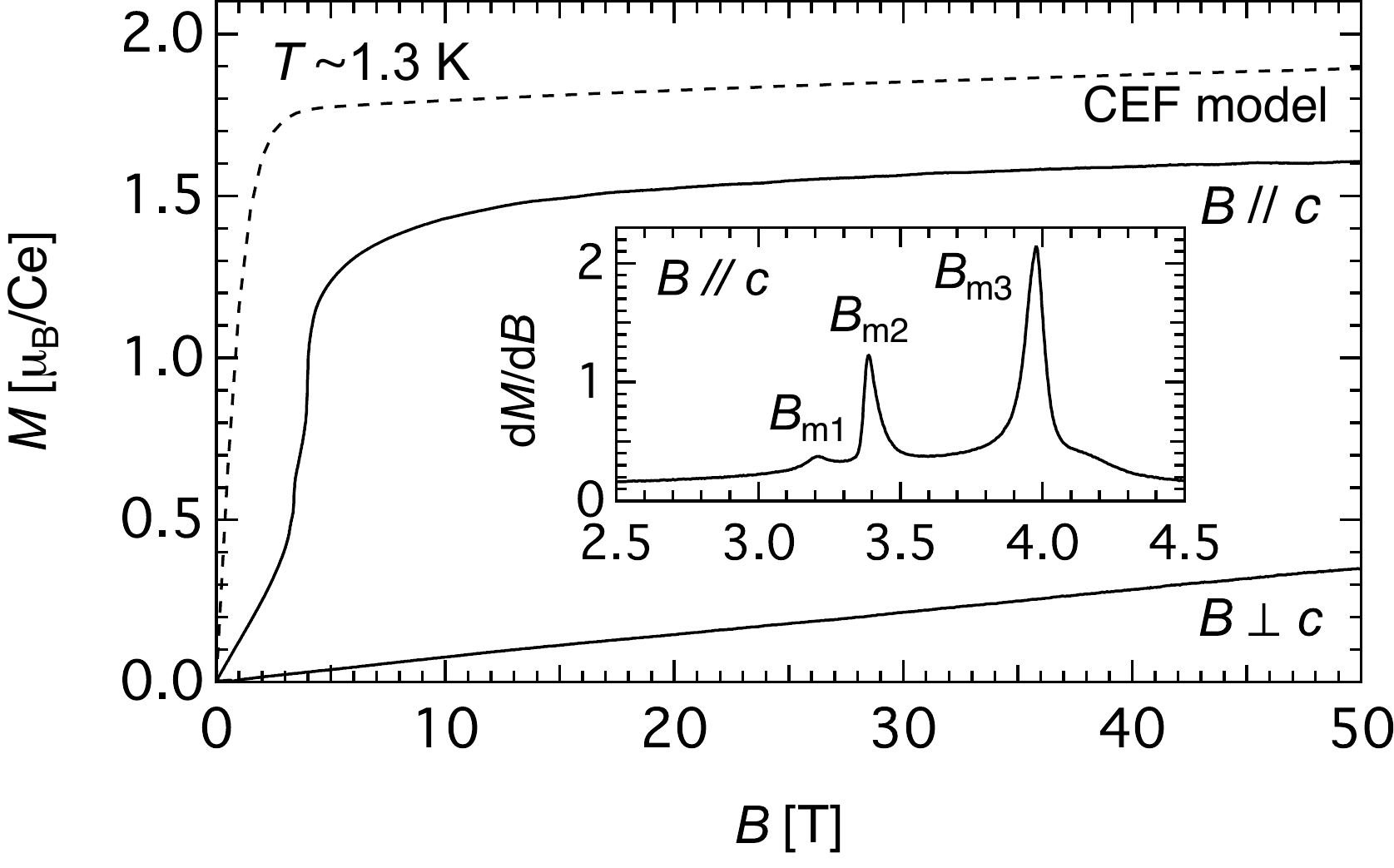}
\caption{\label{fig:CePdAl_M-H_1}
High-field magnetization curves parallel and perpendicular to the $c$-axis of a single crystal CePdAl at 1.3 K. The inset figure shows the differential magnetization along the $c$-axis. The dashed curve is the calculation with the CEF model\cite{Isikawa1996} for $B\parallel c$ assuming $T=1.3$ K.}
\end{figure}
%%%%%%%%%%%%%%%%%%%%%%%%%%%%%%%%%%%%%%%%%%%%%%%%%%%%%%%%%%%%%%%%%%%%%%%%%%%%%%%%%%%%%%%%%%%%%%%%%%%%%%%%%%%%%

The magnetization ($M$) curves of the CePdAl at 1.3 K up to 50 T parallel and perpendicular to the easy axis ($c$-axis) are shown in Fig. \ref{fig:CePdAl_M-H_1}. The inset shows the differential of $M(B)$ curve \textit{i.e.}, d$M(B)$/d$B$ for $B\parallel c$-axis around three metamagnetic transitions. The magnetization along the $c$-axis increases rapidly and shows three metamagnetic transitions\cite{magnetocaloric}. Corresponding to these transitions, d$M(B)$/d$B$ has three clear peaks at $B_{\rm{m}1}$ = 3.2 T, $B_{\rm{m}2}$ = 3.4 T and $B_{\rm{m}3}$ = 4.0 T. In addition, a shoulder-like anomaly is observed at $\sim$4.2 T above $B_{\rm{m}3}$. In the  high magnetic field region, $M(B)$ has a very small slope of $\sim$1.5$\times$10$^{-3}$ $\mu_{\rm{B}}$/T at 45 T and reaches 1.6 $\mu_{\rm{B}}$/Ce at 50 T, smaller than the value expected from the previously determined CEF parameters (dashed curve)\cite{Isikawa1996,CEF_states}. By contrast, the magnetization perpendicular to the $c$-axis shows no anomaly and increases in proportion to $B$ up to 50 T, and the value of the magnetization at 50 T decreases from 1.6 $\mu_{\rm{B}}$/Ce for $B\parallel c$ to 0.35 $\mu_{\rm{B}}$/Ce for $B\bot c$. These results suggest that the large magnetic anisotropy of CePdAl persists even in the strong magnetic field of 50 T.

%%%%%%%%%%%%%%%%%%%%%%%%%%%%%%%%%%%%%%%%%%%%%%%%%%%%%%%%%%%%%%%%%%%%%%%%%%%%%%%%%%%%%%%%%%%%%%%%%%%%%%%%%%%%
\begin{figure}[t]
\includegraphics[width = 3.4in]{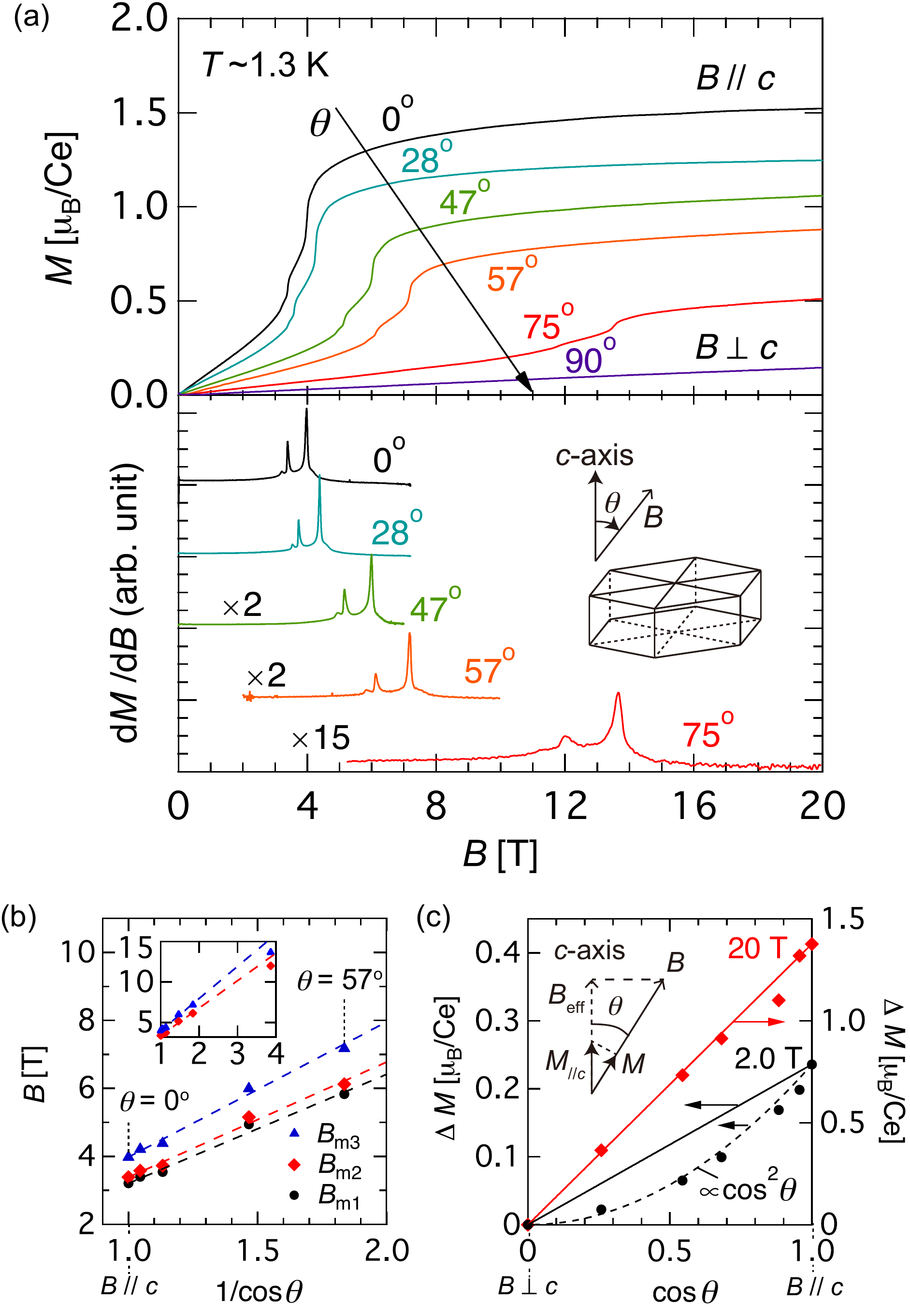}
\caption{\label{fig:CePdAl_M-H_2}
 (Color online)
(a) Magnetization and differential magnetization curves of CePdAl for various crystal orientations. The differential magnetizations at $\theta$ = 47$^\circ$ and 57$^\circ$ are multiplied by 2, and that of $\theta$ = 75$^\circ$ is multiplied by 15 for clarity. 
 (b) Angular dependences of the metamagnetic transition fields $B_{\rm{m}1,2,3}$. The inset shows the wide-area view.
 (c) Angle dependences of the magnetization values below (2.0 T) and above (20 T) the metamagnetic transitions. The constant term $M_0$, which results mainly from the contribution of the CEF exited states, is subtracted as $\Delta M(\theta)=M(\theta)-M_0$. The solid and dashed curves represent the functions in proportional to $\cos\theta$ and $\cos^2\theta$, respectively.}
\end{figure}
%%%%%%%%%%%%%%%%%%%%%%%%%%%%%%%%%%%%%%%%%%%%%%%%%%%%%%%%%%%%%%%%%%%%%%%%%%%%%%%%%%%%%%%%%%%%%%%%%%%%%%%%%%%%%

%%%%%%%%%%%%%%%%%%%%%%%%%%%%%%%%%%%%%%%%%%%%%%%%%%%%%%%%%%%%%%%%%%%%%%%%%%%%%%%%%%%%%%%%%%%%%%%%%%%%%%%%%%%%%
\begin{figure*}[th]
\includegraphics[width = 7in]{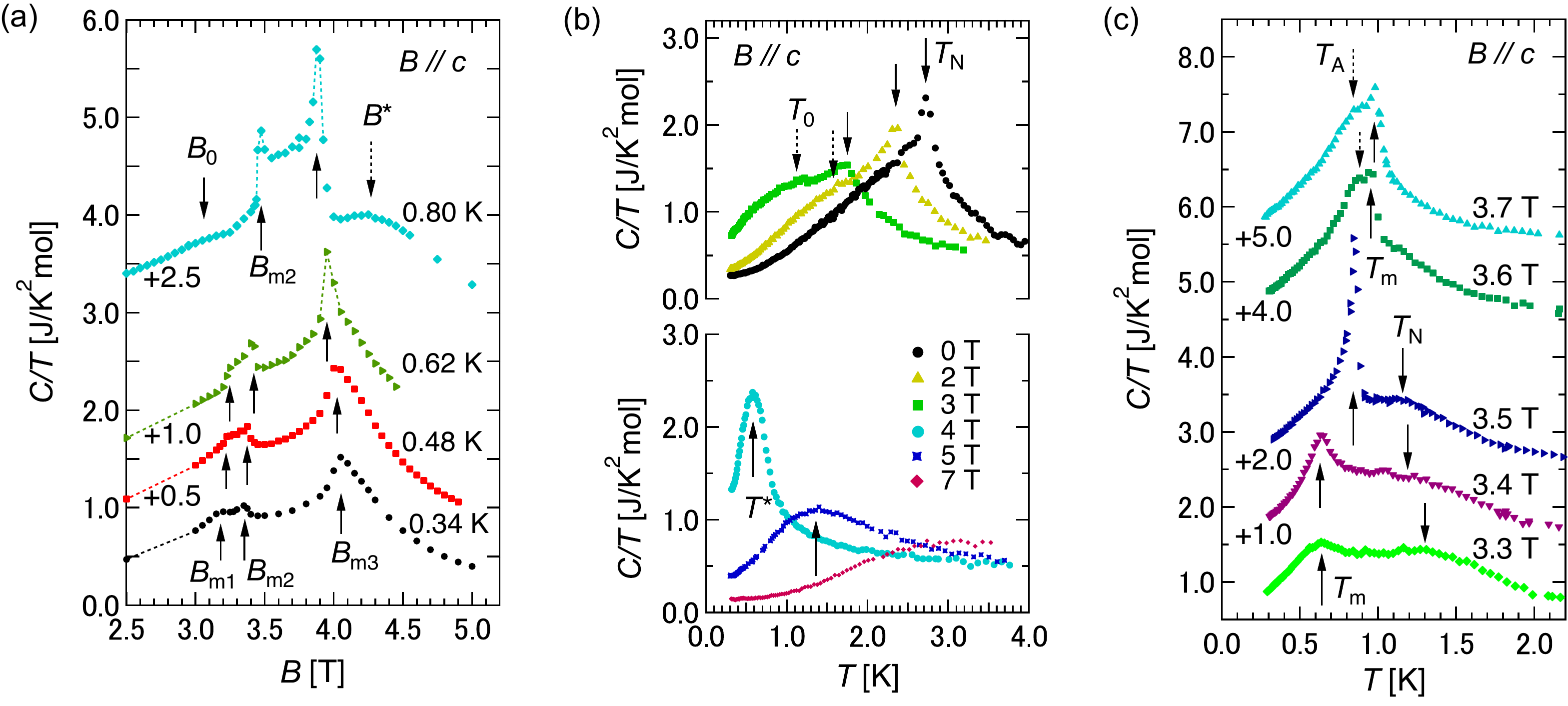}
\caption{\label{fig:HC_CePdAl} 
 (Color online) 
 (a) $C/T$ vs $B$ plots below 1 K. Specific-heat data are offset for clarity. 
The anomalies are distinguished into five magnetic fields of $B_{{\rm m}i}$ ($i=$1,2,3) (up arrows), $B_{0}$ (solid-down arrow), and $B^{\star}$ (dashed-down arrow).
 (b) $C/T$ vs $T$ plots from 0 T to 7 T. The anomalies are distinguished into three temperatures of $T_{\rm{N}}$ (down arrows), $T_{0}$ (dashed-down arrows) and $T^{\star}$ (up arrows). 
 (c) $C/T$ vs $T$ plots in the magnetic fields between 3.0 T and 4.0 T. 
 Specific-heat data are offset for clarity. The sharp anomalies at $T_{\rm{m}}$ are indicated by the up arrows, and the broad anomalies at $T_{\rm{N}}$ and $T_{\rm A}$ are indicated by down arrows and dashed-down arrows, respectively.
}
\end{figure*}
%%%%%%%%%%%%%%%%%%%%%%%%%%%%%%%%%%%%%%%%%%%%%%%%%%%%%%%%%%%%%%%%%%%%%%%%%%%%%%%%%%%%%%%%%%%%%%%%%%%%%%%%%%%%%%

To further investigate the anisotropy of the magnetization and metamagnetic transitions, the angular dependence of $M(B)$ curves are measured [Fig. \ref{fig:CePdAl_M-H_2}(a)]. The angle $\theta$ is measured from the $c$-axis as illustrated in the lower panel of Fig. \ref{fig:CePdAl_M-H_2}(a). The magnetization is gradually suppressed with increasing $\theta$, but is rapidly decreased from $\theta=57^\circ$ to $\theta= 75^\circ$. As seen in the Fig. \ref{fig:CePdAl_M-H_2}(a), each transition field shifts to higher magnetic field, and the anomalies in d$M(B)$/d$B$ become smaller and broader with increasing $\theta$. At $\theta=75^\circ$, the critical fields of the metamagnetic transitions are $B_{\rm{m}2}\sim$12.0 T and $B_{\rm{m}3}\sim$ 13.5 T, whereas the anomaly at $B_{\rm{m}1}$ disappears. The angular dependences of the metamagnetic transition fields are well described by the function of $B_{\rm{m}1,2,3}(\theta)=B_{\rm{m}1,2,3}(\theta = 0)/\cos\theta$ below $\theta=57^\circ$ [Fig. \ref{fig:CePdAl_M-H_2}(b)], indicating that the effective magnetic field along $c$-axis is dominant to the metamagnetic transitions. Thus the high-field states are also Ising-like ordered phases with successive increments of the magnetic moment along $c$-axis, unlike the spin flop phase of conventional antiferromagnetic order. 

%%%%%%%%%%%%%%%%%%%%%%%%%%%%%%%%%%%%%%%%%%%%%%%%%%%%%%%%%%%%%%%%%%%%%%%%%%%%%%%%%%%%%%%%%%

Figure \ref{fig:CePdAl_M-H_2}(c) shows the angular dependences of the magnetization values at 2.0 and 20 T. We subtract the angular-independent term as $\Delta M(\theta) \equiv M(\theta)-M_0$; $M_{\rm 0}= M_{\bot c}$, which is considered to result mainly from the contribution of the CEF excited states. Above $B_{\rm{m}3}$, at 20 T, the magnetization value is well described by the function of $\Delta M_{20\rm{T}}(\theta) = \Delta M_{20\rm{T}}(0)\cos\theta$  [Fig. \ref{fig:CePdAl_M-H_2}(c)]. This ``$\propto$ cos$\theta$" behavior suggests that almost fully-induced magnetic moments on Ce sites are ferromagnetically aligned along the $c$-axis above $B_{\rm{m}3}$, since we detect the magnetization component along magnetic fields as shown in Fig. \ref{fig:CePdAl_M-H_2}(c). The neutron diffraction measurements have also detected the increments of ferromagnetic (100) reflection at the metamagnetic transitions\cite{Prokes2006}.

%%%%%%%%%%%%%%%%%%%%%%%%%%%%%%%%%%%%%%%%%%%%%%%%%%%%%%%%%%%%%%%%%%%%%%%%%%%%%%%%%%%%%%%%%%

In contrast, the magnetization at 2.0 T below $B_{\rm{m}1}$ is proportional to cos$^2\theta$ rather than cos$\theta$ [Fig. \ref{fig:CePdAl_M-H_2}(c)]. This is reasonable if the magnetic moments along $c$-axis $M_{\parallel c}$ are induced by the effective magnetic field $B_{\rm{eff}}= B\cos\theta$ as $M_{\parallel c} = \chi_{\parallel c}B_{\rm{eff}} = \chi_{\parallel c}B\cos\theta$. In this case, we detect the magnetization component along the applied field $M = M_{\parallel c} \cos\theta = \chi_{\parallel c} B\cos^2\theta$ [Fig. \ref{fig:CePdAl_M-H_2}(c)]. The moments on the ordered Ce(1) and Ce(3) sites may hardly contribute to the observed magnetization below $B_{\rm{m}1}$, since they cannot rotate freely due to the Ising-type anisotropy. We would like to point out that the partially disordered Ce(2) sites may contribute to such cos$^2\theta$ dependence due to appearance of paramagnetic moments on Ce(2) sites, which are induced by the breaking of Kondo screening with increasing field. The presence of field-induced paramagnetic moments on Ce(2) sites might  be supported by the fact that the magnetization at $\sim B_{\rm{m}1}$ reaches one-third of the full moment $M_{\rm 50 T}$.

%%%%%%%%%%%%%%%%%%%%%%%%%%%%%%%%%%%%%%%%%%%%%%%%%%%%%%%%%%%%%%%%%%%%%%%%%%%%%%%%%%%%%%%%%%
%
%\subsection{\label{sec:level31}Specific heat}

Next we report the results of the detailed measurements of the specific heat ($C$) around the metamagnetic transitions. We first show the magnetic-field dependences of specific heat for $B\parallel c$-axis below 1 K in Fig. \ref{fig:HC_CePdAl}(a). At  0.34 K, we observe two kink anomalies at $B_{\rm{m1}}=3.2$ T and $B_{\rm{m2}}=$ 3.4 T as well as a maximum at $B_{\rm{m3}}=$ 4.0 T, corresponding to the three metamagnetic transitions. They become sharper with increasing temperature up to 0.8 K. At $B_{\rm{m3}}$,  the magnetization jump becomes largest (Fig. \ref{fig:CePdAl_M-H_1}), accompanying the maximum in $C/T$.  At 0.80 K, the two peaks at $B_{\rm{m1}}$ and $B_{\rm{m2}}$ merge into one peak. Moreover, two additional broad peaks appear  at  $B_0 < B_{\rm{m1}}$ and $B^{\star} > B_{\rm{m3}}$, indicating existence of crossovers below $B_{\rm{m1}}$ and in the paramagnetic region above $B_{\rm{m3}}$.

%%%%%%%%%%%%%%%%%%%%%%%%%%%%%%%%%%%%%%%%%%%%%%%%%%%%%%%%%%%%%%%%%%%%%%%%%%%%%%%%%%%%%%%%%%

Figure \ref{fig:HC_CePdAl}(b) show the temperature dependences of the specific heat at zero and magnetic fields up to 7 T for $B\parallel c$-axis. $C/T$ shows the different behavior between below ($B\leq$ 3 T) and above ($B\geq$ 4 T) metamagnetic transitions. A large peak anomaly in $C(T)/T$ at 0 T is due to the antiferromagnetic order ($T_{\rm{N}}$ = 2.7 K). With increasing field up to 3 T, the peak at $T_{\rm{N}}$ becomes smaller.  In addition, a broad anomaly appears around $T_0 (<T_{\rm{N}})$ at 2 and 3 T. On the other hand, in high fields above 4 T, a broad and large Schottky-like anomaly observed at $T^{\star}$, which occur in paramagnetic phase above metamagnetic transition fields. This anomaly shifts to higher temperature and becomes broader with increasing field.

%%%%%%%%%%%%%%%%%%%%%%%%%%%%%%%%%%%%%%%%%%%%%%%%%%%%%%%%%%%%%%%%%%%%%%%%%%%%%%%%%%%%%%%%%%

\begin{figure}[t]
\includegraphics[width = 3.5in]{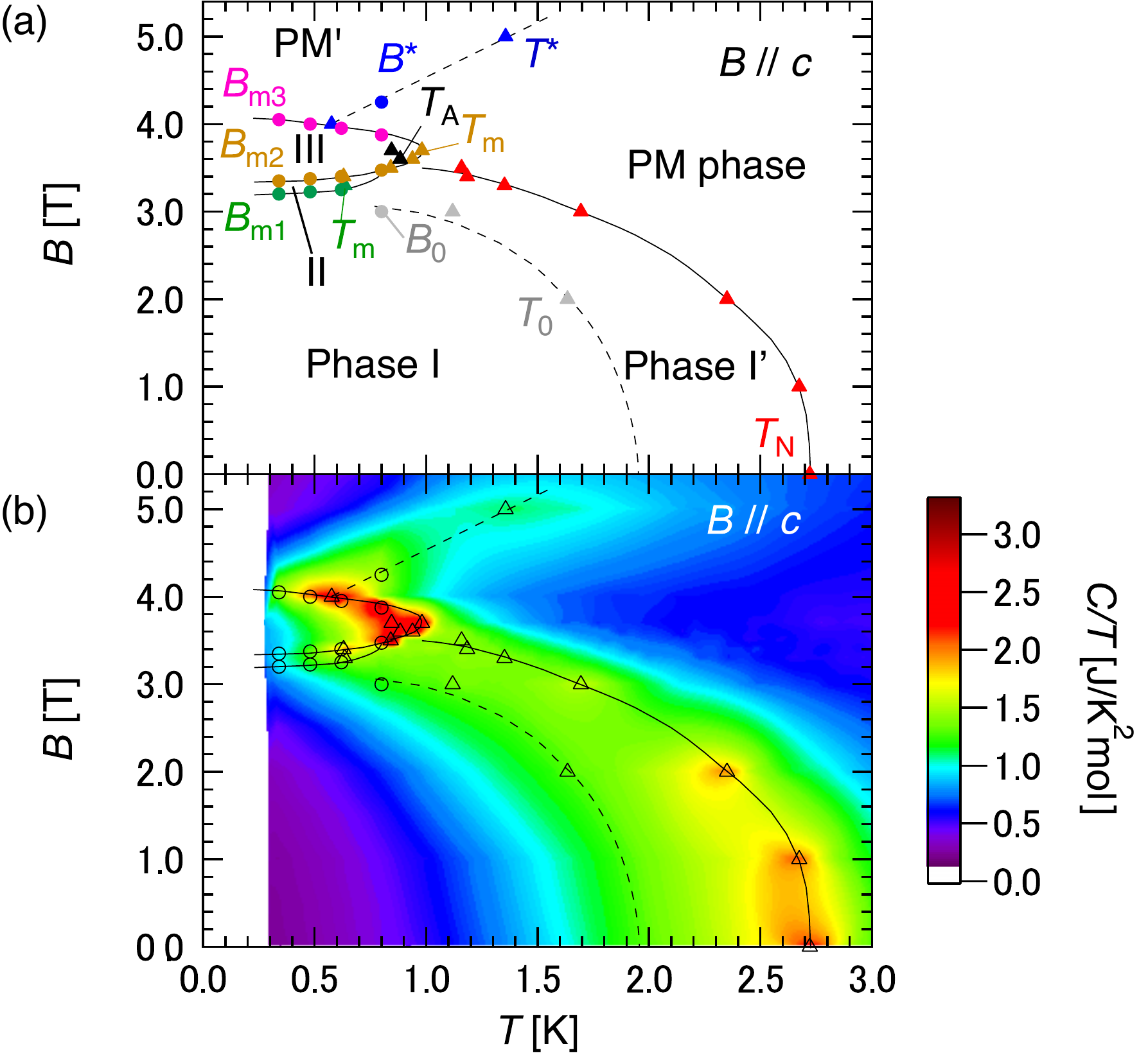}
\caption{\label{fig:phase_diagram} 
  (Color online) 
$B$-$T$ phase diagram of CePdAl ($B\parallel c$). The triangles and the circles indicate the critical temperatures and magnetic fields, respectively. (b) $C(T,B)/T$ mapped on the $B$-$T$ phase diagram of CePdAl. The color changes from violet to dark red with increasing $C/T$.}
\end{figure}

Figure \ref{fig:HC_CePdAl}(c) shows the $C(T)/T$ vs $T$ plots around the three metamagnetic transitions. Sharp ($T_{\rm{m}}$) and broad ($T_{\rm{N}}, T_{\rm A}$) anomalies are very sensitive to the magnetic fields, and show different field dependences. The anomaly of the antiferromagnetic transition at $T_{\rm{N}}$ broadens with increasing magnetic field, and remains up to 3.5 T. At $T_{\rm{m}}(<T_{\rm{N}})$, the sharp anomaly appears in fields between 3.3 to 3.7 T, which seems to be related to the metamagnetic transitions as discussed later with obtained phase diagram [Fig. \ref{fig:phase_diagram}(a)]. The anomaly becomes huge between 3.5 T and 3.7 T, while $T_{\rm{N}}(B)$ vanishes around 3.6 T. Moreover, another anomaly appears at $T_{\rm A}(<T_{\rm{m}})$, indicating the existence of the crossover below $T_{\rm{m}}$.

%%%%%%%%%%%%%%%%%%%%%%%%%%%%%%%%%%%%%%%%%%%%%%%%%%%%%%%%%%%%%%%%%%%%%%%%%%%%%%%%%%%%%%%%%%

Figure  \ref{fig:phase_diagram}(a) shows the $B$-$T$ phase diagram of CePdAl for the magnetization-easy axis  ($B\parallel c$) obtained from the present specific-heat measurements. The solid and dashed lines indicate the phase boundary and crossover, respectively. The phase diagram consists of three ordered phases I-III and a paramagnetic phase. The phase boundaries between I-III and PM' are accompanied by the metamagnetic transitions. The antiferromagnetic ordered phase I is divided by a crossover line $T_0(B)$ into two regions, where the higher $T$ region is defined as I'. On the other hand, the paramagnetic phase is separated into PM and PM' regions by the broad Schottky-like anomaly in $C(T)/T$ above 4 T. Moreover, inside the phase III, there is a broad hump anomaly at $T_{\rm A}$ below $T_{\rm{m}}$.

%%%%%%%%%%%%%%%%%%%%%%%%%%%%%%%%%%%%%%%%%%%%%%%%%%%%%%%%%%%%%%%%%%%%%%%%%%%%%%%%%%%%%%%%%%%%%%%%%%%%%%%%%%%%%%%
%%%%%%%%%%%%%%%%%%%%%%%%%%%%%%%%%%%%%%%%%%%%%%%%%%%%%%%%%%%%%%%%%%%%%%%%%%%%%%%%%%%%%%%%%%%%%%%%%%%%%%%%%%%%%%%

It seems that these metamagnetic phase boundaries between I-III and PM' are merged into one boundary. We have observed hysteresis behavior at $B_{{\rm m}i}$ ($i=$1,2,3) from dc magnetization measurements, which will be reported elsewhere. Thus the merging point is possibly a tri-critical point, at which the line of second-order phase transition at $T_{\rm{N}}(B)$ meet the line of first-order phase transition, \textit{i.e.}, the boundary of phase III. Around the critical point, there exist a strong enhancement of the fluctuation. We shall see the contour plot of $C(T,B)/T$ of CePdAl [Fig. \ref{fig:phase_diagram}(b)] in order to visualize the effect of  magnetic fluctuation, mapping on the obtained $B$-$T$ phase diagram. Owing to the suppression of the antiferromagnetic transition by the magnetic field, the green region of the large $C/T$ shifts to low temperature below 4 T. The dark-red region, indicating very large $C/T$ spreads around the meeting point of the phase boundary lines $B_{\rm m2}(T)$ and $T_{\rm N}(B)$. In particular, $C/T$ is strongly enhanced, showing  a $\lambda$-type anomaly at 3.5 T and 0.84 K through the phase transition between the III and PM phases. This strong enhancement of $C/T$ also supports the possible presence of a critical point around 3.5 T.

%%%%%%%%%%%%%%%%%%%%%%%%%%%%%%%%%%%%%%%%%%%%%%%%%%%%%%%%%%%%%%%%%%%%%%%%%%%%%%%%%%%%%%%%%%%%%%%%%%%%%%%%%%%%%%%%
\begin{figure}[tp]
\includegraphics[width = 3.4in]{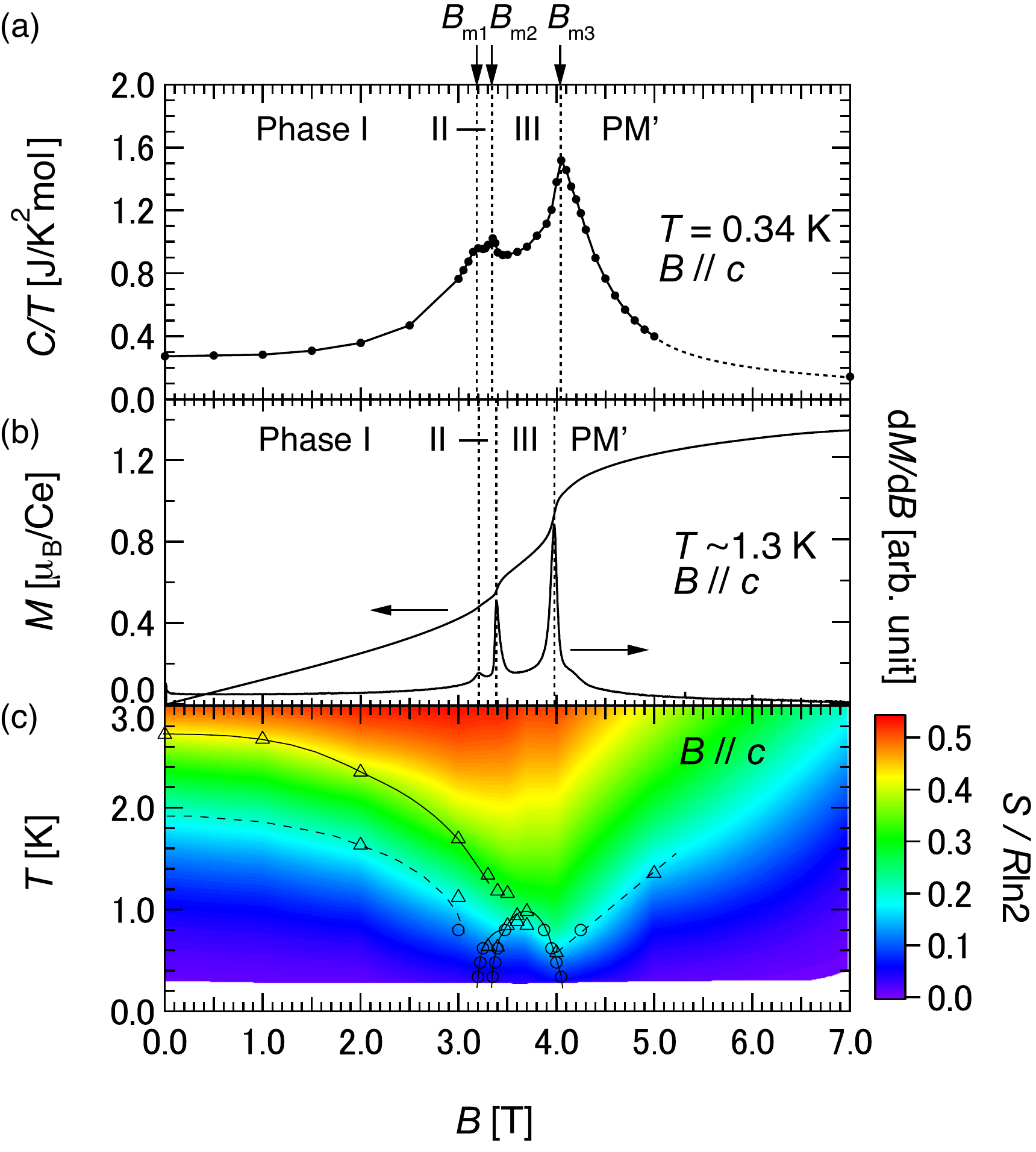}
\caption{\label{fig:gamma_mag} 
  (Color online)  
Field dependences of  
(a) the specific heat ($C$) at 0.34 K (dashed curve between 5 and 7 T is a guide to the eye),  
 (b) magnetization curve with its derivative up to 7 T, and 
   (c) entropy $S/R$ln2 mapped on the $T$-$B$ phase diagram.
 The dashed lines in Fig. \ref{fig:gamma_mag}(a) and (b) indicate the metamagnetic transition fields.
%%%%% 
The solid and dashed lines in Fig. \ref{fig:gamma_mag}(c) indicate the lines of phase transition and crossover, respectively.
}
\end{figure}
%%%%%%%%%%%%%%%%%%%%%%%%%%%%%%%%%%%%%%%%%%%%%%%%%%%%%%%%%%%%%%%%%%%%%%%%%%%%%%%%%%%%%%%%%%%%%%%%%%%%%%%%%%%%%%%%

Let us discuss the relationship between the obtained magnetic phase diagram and heavy-fermion state in CePdAl. Figures \ref{fig:gamma_mag}(a) and (b) show magnetic-field dependence of $C(B)/T$ at the lowest temperature 0.34 K and the magnetization curve at 1.3 K. It is stressed that there exist disordered  $f$ electrons even in the antiferromagnetic state, as revealed by the previous NMR studies \cite{Oyamada2008}. In the phase I, \textit{i.e.}, the low-field antiferromagnetic phase, $C/T$ increases towards $B_{\rm m1}$, and it becomes $\sim$0.96 JK$^{-2}$mol$^{-1}$ at 3 T. The large $C/T$ value remains in the field-induced phases II and III. The phase transition from phase III to the paramagnetic phase occurs accompanying with a large maximum of $C/T \sim $1.5 JK$^{-2}$mol$^{-1}$. This enhancement of $C/T$ is probably related to the density of states of the heavy electrons as well as magnetic fluctuations in CePdAl. Very recently, it has been reported  that the $A$ coefficient of the resistivity [$\rho(T)  =  \rho_{0} + AT^2$] of CePdAl reaches a large maximum value of $A = $ 12 $\mu \Omega$cmK$^{-2}$ just above $B_{\rm{m3}}$\cite{Zhao2016}. These large values of $C/T$ [Fig. \ref{fig:gamma_mag}(a)] and $A$ coefficient, according to the Kodowaki-Wood's relation\cite{Kadowaki1986}, indicate that $f$ electrons in CePdAl form a strongly correlated heavy-fermion state, as in the  heavy-fermion superconductor CeCu$_{2}$Si$_{2}$ \cite{Steglich1979}.

%%%%%%%%%%%%%%%%%%%%%%%%%%%%%%%%%%%%%%%%%%%%%%%%%%%%%%%%%%%%%%%%%%%%%%%%%%%%%%%%%%%%%%%%%%

The essential point for understanding  the low-$T$ physics of CePdAl is to explain how a $f$ electron releases the entropy of $R$ln2 for the Kramers doublet as the localized CEF ground state. Figure. \ref{fig:gamma_mag}(c) shows the entropy of CePdAl obtained from the present specific-heat measurements. The contribution of the lattice is subtracted by using the specific heat of YPdAl\cite{Kitazawa1994}. At zero magnetic field, the entropy above $T_{\mathrm{N} }$ is $\sim$0.5$R$ln2, implying that the entropy of the $f$ electron is already reduced at low temperatures due to the Kondo effect. The entropy of the $f$ electron is further released by the incommensurate magnetic ordering, but the entropy release is not  sufficient  probably due to the geometrical frustration. Then, the partially Kondo screening for the Ce(2) site occurs at lower temperature below $T_{\mathrm{N}}$. The crossover line of $T_{0}(B)$, represented by a dashed line, might be a signature of the partial Kondo screening, since the remaining entropy of $\sim0.3R$ln2 seems to be  released at $T_{0}$. This crossover might also be related to the fixing of incommensurate propagation vector component $\tau$\cite{Prokes2006}. In magnetic field, the partial Kondo screening begins to break with a magnetic field of several Tesla, since we also estimate the Kondo temperature of several Kelvin by using the entropy value at 0 T, \textit{i.e.},  $T_{\rm K}\sim 2T(S=0.5R\ln 2)\sim 6$ K. Interestingly, the entropy as a function of $B$ in paramagnetic phase shows a maximum around 3-4 T (red region) [Fig. \ref{fig:gamma_mag}(c)]. This behavior suggests that the breaking of the Kondo effect in the magnetic field induces the geometrical frustration again. Thus, to lift the frustration, magnetic phases II and III appear as the field-induced ground states between 3 and 4 T.

%%%%%%%%%%%%%%%%%%%%%%%%%%%%%%%%%%%%%%%%%%%%%%%%%%%%%%%%%%%%%%%%%%%%%%%%%%%%%%%%%%%%%%%%%%

In the field-induced paramagnetic phase  above  $B_{\rm{m3}}$, the large value $C(B)/T$ dramatically decreases  with increasing field, suggesting the reduction of  its heavy-effective mass for $B> B_{\rm{m3}}$. Simultaneously,   the magnetization  $M(B)$ does not reach the full moment immediately, with a hump structure for  its derivation d$M(B)$/d$B$  in a field range of $4.0 $  $<$ $B$ $<$ 4.2 T [Fig. \ref{fig:gamma_mag}(b)]. Such behaviors are possibly related to  the effects of geometrical frustration as well as the remaining of the Kondo effect. On the other hand, in high-field region above $\sim$20 T, the slope of the magnetization curve is roughly described by the CEF model (Fig. \ref{fig:CePdAl_M-H_1}). Moreover, above  4 T, a broad Schottky-like anomaly appears in $C/T$, probably due to the Zeeman splitting of the CEF ground state. These behaviors suggest that the heavy quasiparticles  dressed by the Kondo cloud in the low-field region gradually recover a well localized behavior with increasing field.

%%%%%%%%%%%%%%%%%%%%%%%%%%%%%%%%%%%%%%%%%%%%%%%%%%%%%%%%%%%%%%%%%%%%%%%%%%%%%%%%%%%%%%%%%%

It is also interesting to discuss the physical properties around the  phase boundary $B_{\rm{m3}}(T)$ as $T \rightarrow$ 0 K. As seen in Fig. \ref{fig:phase_diagram}(b), the $C(T,B)/T$ is not enhanced  towards zero temperature around $B_{\rm{m3}}$. It is thus considered that the line of $B_{\rm{m3}}(T)$ remains first-order, unlike the presence of a quantum critical point, where a second-order phase transition line vanishes as $T \rightarrow$ 0 K. The situation of the quantum phase transition  at $B_{\rm{m3}}$ in CePdAl is quite different from that of another heavy-fermion antiferromagnet YbRh$_{2}$Si$_{2}$ which shows the field-induced quantum criticality\cite{Tokiwa2009}. Here, it has been reported  that non-Fermi-liquid  behaviors near  a quantum critical point for a two-dimensional  antiferromagnetic system are induced by pressure \cite{Akamaru2002,Goto2002} and Ni substitution, \textit{i.e.},  CePd$_{1-x}$Ni$_{x}$Al\cite{Fritsch2014, Isikawa2000}. Our results suggest that magnetic field is not a tuning parameter which induces  the quantum critical point in this system. In pure CePdAl, non-Fermi-liquid behavior has not been observed around  $B_{\rm{m3}} (T)$ from low-$T$ resistivity measurements\cite{Zhao2016}.

%%%%%%%%%%%%%%%%%%%%%%%%%%%%%%%%%%%%%%%%%%%%%%%%%%%%%%%%%%%%%%%%%%%%%%%%%%%%%%%%%%%%%%%%%%

We finally discuss a possibility of  a spin-liquid phase around $B_{\rm{m3}}$, as reported by recent resistivity studies\cite{Zhao2016}. From our specific-heat measurements,  a broad anomaly  is observed at $T_{\rm A}$ just below the field-induced transition temperature  $T_{\rm{m}}(B)$ [Fig. \ref{fig:HC_CePdAl} (c)]. This  anomaly ($T_{\rm A}$) inside the magnetic ordered phase III may be distinguished from  the behavior of the spin-liquid phase, which is \textit{not}  characterized by any ordering. Thus we have not yet obtained any thermodynamic evidence for a spin-liquid phase from  specific-heat data. Nevertheless, the hump structure in the derivative magnetization  d$M(B)$/d$B$ above $B_{\rm{m3}}$ might be related to the reported resistivity anomaly. Further studies from other probes are intriguing  to  gain more insight into the novel physical properties around the field-induced phases in CePdAl.

%%%%%%%%%%%%%%%%%%%%%%%%%%%%%%%%%%%%%%%%%%%%%%%%%%%
\section*{\label{sec:level5}Conclusions}
%%%%%%%%%%%%%%%%%%%%%%%%%%%%%%%%%%%%%%%%%%%%%%%%%%%

To conclude, thermodynamic properties of  CePdAl have been studied by means of dc pulsed-field magnetization and low-$T$ specific-heat measurements using a single crystalline sample. In low-field antiferromagnetic phase I, the crossover anomaly at $T_{0}(B)$, as  observed in $C(T)/T$, may indicate an occurrence  of the partial Kondo screening below $T_{\mathrm{N}}$, releasing the residual entropy of $0.3R$ln2. With increasing magnetic field, this screening is gradually broken, and as a result, the paramagnetic moment with the one-third of the full moment $M_{\rm 50 T}$ appears around $B_{\rm m1}$. To lift the frustration caused by the appearance of the magnetic moment on the partial disorder sites in high-field region, the magnetic phases II and III appear as the ground state. The magnetization measurements demonstrate that these high-field states II and III are strongly Ising-type ordered phases. Moreover,  the  large enhancement of $C/T$ at 3.5-3.7 T may imply that the strong magnetic fluctuation exists around the tri-critical point. With increasing field above $B_{\rm m3}$, the heavy-fermion state, accompanying with the partial disorders, gradually breaks down above 4 T, and $f$ electrons recover the localized behavior, which can be described by the CEF model. These thermodynamic results along with the obtained unusual magnetic phase diagram deepen the understanding for the low-$T$ ground state of this geometrically frustrated Ising-type quasi-Kagome Kondo lattice.

\section*{\label{sec:level6}Acknowledgements}
We would like to thank M. Imada, Y. Yamaji, and S. Kambe for valuable discussions and comments. K. M. was supported by Japan Society for the Promotion of Science through Program for Leading Graduate Schools (MERIT).

%%%%%%%%%


\begin{thebibliography}{99}
\bibitem{magnetocaloric} The temperature during the measurements can be decreased below 1 K due to the large magnetocaloric effect of the antiferromagnet in the magnetic field.
\bibitem{CEF_states} The CEF ground state is represented as the wavefunction of $0.89|\pm 5/2\rangle + 0.15|\mp 3/2\rangle + 0.43|\pm 1/2\rangle$, and the energies of the first and second excited states are assumed to be $\Delta_1\sim170$ K and $\Delta_2\sim410$ K, respectively.
\end{thebibliography}
\end{document}